\documentclass[]{emulateapj}

\newcommand{\wise}{\textit{WISE}}

\def\spose#1{\hbox to 0pt{#1\hss}}
\def\simlt{\mathrel{\spose{\lower 3pt\hbox{$\mathchar"218$}}
     \raise 2.0pt\hbox{$\mathchar"13C$}}}
\def\simgt{\mathrel{\spose{\lower 3pt\hbox{$\mathchar"218$}}
     \raise 2.0pt\hbox{$\mathchar"13E$}}}

\usepackage{amsmath}
\usepackage{graphicx}
\usepackage{lpic}
\usepackage{mathtools}

\shorttitle{Angular Clustering of WISE-Selected AGN}
\shortauthors{Donoso et al.}

\begin{document}
\title{The Angular Clustering of WISE-Selected AGN:\\  Different Haloes for Obscured and Unobscured AGN}
\author{E. Donoso$^{1,2}$, Lin Yan$^3$, D. Stern$^4$, R. J. Assef$^{4,5,6}$}
\affil{$^1$Instituto de Ciencias Astron\'{o}micas, de la Tierra, y del Espacio (ICATE), 
5400, San Juan, Argentina}
\affil{$^2$Spitzer Science Center, California Institute of Technology, Pasadena, CA 91125, USA}
\affil{$^3$Infrared Processing and Analysis Center, Department of Astronomy, California Institute of Technology, Pasadena, CA 91125, USA}
\affil{$^4$Jet Propulsion Laboratory, California Institute of Technology,
Pasadena, CA 91109, USA}
\affil{$^5$NASA Postdoctoral Program Fellow}
\affil{$^6$N\'{u}cleo de Astronom\'{i}a de la Facultad de Ingenier\'{i}a, Universidad Diego Portales, Av. Ej\'{e}rcito 441, Santiago, Chile}

\begin{abstract}
We calculate the angular correlation function for a sample of
$\sim$170,000 AGN extracted from the {\it Wide-field Infrared Survey
Explorer} (\wise) catalog, selected to have red mid-IR colors ($W1
- W2 > 0.8$) and 4.6~$\mu$m flux densities brighter than 0.14~mJy). 
The sample is expected to be $> 90\%$ reliable at identifying AGN, 
and to have a mean redshift of $\langle z \rangle=1.1$. In total, 
the angular clustering of \wise\ AGN is
roughly similar to that of optical AGN. We cross-match these objects
with the photometric SDSS catalog and distinguish obscured sources
with $r - W2 > 6$ from bluer, unobscured AGN. Obscured sources
present a higher clustering signal than unobscured sources. Since
the host galaxy morphologies of obscured AGN are not
typical red sequence elliptical galaxies and show disks in many cases, 
it is unlikely that the increased clustering
strength of the obscured population is driven by a host galaxy
segregation bias. By using relatively complete redshift distributions
from the COSMOS survey, we find obscured sources at $\langle z
\rangle \sim 0.9$ have a bias of $b = 2.9 \pm 0.6$ and are hosted
in dark matter halos with a typical mass of $\log(M/M_\odot\,
h^{-1})\sim 13.5$. In contrast, unobscured AGN at $\langle z \rangle
\sim 1.1$ have a bias of $b = 1.6 \pm 0.6$ and inhabit halos of
$\log(M/M_\odot\, h^{-1}) \sim 12.4$.  These findings suggest that
obscured AGN inhabit denser environments than unobscured AGN, and
are difficult to reconcile with the simplest AGN unification models,
where obscuration is driven solely by orientation.
\end{abstract}


\section{Introduction}
In recent years, a large body of evidence suggests that the evolution
and properties of active galactic nuclei (AGN) are tightly linked
not only to the properties of their hosting galaxies, but also to
the environment that these host galaxies inhabit. The most clear
example of this perhaps comes from radio-loud AGN, which have long
been known to be primarily hosted by giant, massive, elliptical
galaxies, which are predominantly found in very dense environments
(\citealt{matthews}; \citealt{best}; \citealt{donoso};
\citealt{wylezalek}). In general, X-ray AGN have also been found
to be strongly clustered (\citealt{gilli05}; \citealt{georgakakis};
\citealt{coil}), though X-ray AGN out to $z \sim 1$ with harder
X-ray spectra, e.g., type-2, or obscured X-ray AGN, are preferentially
found in underdense regions (\citealt{tasse}).

Large redshift surveys such as the Sloan Digital Sky Survey (SDSS,
\citealt{york}) and the 2dF QSO Redshift survey (\citealt{croom04})
have enabled detailed studies of optical quasars, and have shown
that their clustering was larger in the past in such a way that
optically selected quasars seem to be hosted by halos of roughly
constant mass, a few times $10^{12}\, M _\odot$, out to $z \sim 3-4$.

The advent of the {\it Spitzer Space Telescope} opened a new,
mid-infrared (mid-IR) window to AGN populations, providing samples
that are relatively insensitive to the dust extinction that affects
quasar surveys in the optical, ultraviolet (UV) and soft X-ray ($<
10$~keV) bands.  \citet{stern05} developed a simple selection
technique based on IRAC colors that identifies luminous AGN essentially
independent of their obscuration, and thus is particularly useful
for identifying the dominant population of obscured AGN that were
largely missed in previous surveys (see also \citealt{lacy};
\citealt{donley}). However, it is the recent launch of the {\it
Wide-field Infrared Survey Explorer} (\wise; \citealt{wright}) that
has made it possible to efficiently and robustly recover AGN over
the entire sky, including both unobscured and obscured sources.

The most widely accepted idea about the physical origin of obscuration
is the presence of a thick dust torus that, when viewed sideways,
blocks the central part of the AGN and hides many of the quasar-like
features observed in unobscured AGN (\citealt{antonucci}; \citealt{urry}).
The first indirect evidence in favor of a torus was the detection
of polarized broad emission lines, characteristic of unobscured AGN,
in a fraction of well known obscured AGN due to the scattering toward
the line of sight by free electrons just above (or below) the torus
(see \citealt{heisler}). As an alternative to orientation-driven
or torus models of AGN obscuration, it is also plausible that at
least part of obscuration could be caused by the interstellar medium
(ISM) of the host galaxy or by larger, $\sim$kpc-scale clouds of
cool dust (e.g., \citealt{martinezs}). Specifically, galaxy formation
simulations by \citet{hopkins08} predict enhanced AGN activity after
galaxy mergers, which is initially obscured by kpc-scale dust clouds
but is later laid bare as AGN feedback pushes out the obscuring
material.

A basic prediction of the orientation-driven AGN unification models
is that similarly selected AGN should populate similar environments.
While some differences are clearly evident based on intrinsic AGN 
luminosity or radio-loudness (e.g., \citealt{donoso}; \citealt{falder}),
the expectation is that obscured (or type-2) and unobscured (or type-1)
AGN of similar luminosity
and radio power should reside in similar environments. However,
relatively little is known about the clustering of obscured AGN,
particularly those identified at mid-IR wavelengths. \citet{gilli09}
studied the spatial clustering of X-ray AGN at $z \sim 1$, finding
no significant difference in clustering strength between obscured
and unobscured X-ray selected AGN. Similarly, from a matched sample
of powerful radio-loud AGN at $1 < z < 3$, \citet{wylezalek} found
that radio-loud quasars (e.g., unobscured radio-loud AGN) reside
in similar environments to high-redshift radio galaxies (e.g.,
obscured radio-loud AGN). In contrast, \citet{hickox} analyzed a
sample of 806 {\it Spitzer} mid-IR-selected quasars at $0.7 < z <
1.8$ in the Bo\"{o}tes field. They found marginal ($< 2\sigma$)
evidence that obscured quasars have a larger bias and populate more
massive dark matter halos.

These studies, while powerful due to the availability of spectroscopic
redshifts and/or a large number of photometric bands, suffer the
typical limitations of deep pencil-beam surveys, providing samples of a
few hundred objects at most. In this paper we adopt a complementary
approach by combining the \wise\ and SDSS data sets over thousands
of square degrees. We select AGN based on the \wise\ 3.4~$\mu$m
($W1$) and 4.6~$\mu$m ($W2$) bands using selection criteria recently
developed and demonstrated by \citet{stern2012} and \citet{assef2013}.
To quantify the clustering, we undertake a correlation analysis,
which is arguably the most powerful method for studying the
distribution of galaxies. The angular correlation function measures
the projected clustering of galaxies by comparing the distribution
of galaxy pairs relative to that of a random distribution. While a
less direct probe than the spatial correlation function, $\xi(r)$,
the angular correlation function is a powerful approach as it
can be applied to wide-area surveys and large samples of galaxies,
overcoming the limitations of small number statistics and cosmic
variance. In this work we focus on the angular correlation of AGN.
Adopting a preliminary estimate of the redshift distribution of
\wise-selected AGN, we derive the absolute bias and estimate the
typical mass of the dark matter halos that host them.

This paper is organized as follows. In \S2 we describe the surveys
used in this work. In \S3 we describe mid-IR selection of AGN using
the \wise\ survey and detail the colors, morphologies, and redshift
distribution of such sources. Section~4 presents the angular
clustering measurements, \S5 presents the results and conclusions,
and \S6 summarizes these results and discusses the implications of
this work.

Throughout the paper we assume a flat concordance $\Lambda$CDM
cosmology, with $\Omega_{m}=0.3$, $\Omega_{\Lambda}=0.7$, and $H_0
= 70\, {\rm km}\, {\rm s}^{-1}\, {\rm Mpc}^{-1}$.  Unless otherwise
noted, all magnitudes in this paper refer to the Vega system.

\begin{figure}
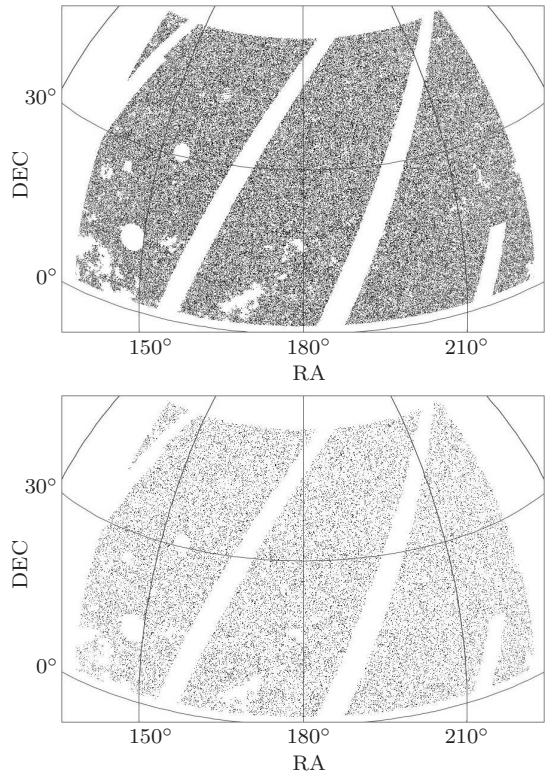
  
\begin{lpic}[l(12mm),b(8mm),clean]{agn_v1(0.3)}
\lbl[c]{40,-6;$150^\circ$}
\lbl[c]{110,-6;$180^\circ$}
\lbl[c]{180,-6;$210^\circ$}
\lbl[r]{-1,25;$0^\circ$}
\lbl[r]{-1,105;$30^\circ$}
\lbl[c]{110,-18;RA}
\lbl[c]{-18,70,90;DEC}
\end{lpic}
\begin{lpic}[l(12mm),b(7mm),clean]{agnnopt_v1(0.3)}
\lbl[c]{40,-6;$150^\circ$}
\lbl[c]{110,-6;$180^\circ$}
\lbl[c]{180,-6;$210^\circ$}
\lbl[r]{-1,25;$0^\circ$}
\lbl[r]{-1,105;$30^\circ$}
\lbl[c]{110,-18;RA}
\lbl[c]{-18,70,90;DEC}
\end{lpic}
\caption{Equatorial coordinates of \wise\ AGN projected onto the
celestial sphere after the masking procedure described in
\S\ref{sec:mask} to remove areas with data of compromised quality
(i.e., around Moon trails, large sources, bright stars, and areas of high Galactic
absorption). The top panel shows all AGN candidates with $W1 - W2
> 0.8$ and $W2 < 15.05$, while the bottom panel shows \wise\ AGN
lacking optical counterparts in SDSS. No obvious large-scale
differences are evident, suggesting the latter are not related to
Galactic sources, extinction or image artifacts.}
\label{fig:mask}
\end{figure}

\section{Data}
\subsection{Wide-field Infrared Survey Explorer}
The \wise\ satellite mapped the full sky in four bands centered at
3.4, 4.6, 12 and 22~$\mu$m (bands $W1$, $W2$, $W3$ and $W4$, respectively),
achieving 5-$\sigma$ point source sensitivities better than 0.08,
0.11, 1 and 6~mJy, respectively. Every part of the sky has been
observed typically $\sim 10$ times, except near the ecliptic poles
where the coverage is much higher. Astrometric precision is better
than 0\farcs15 for high signal-to-noise (SNR) sources (\citealt{jarrett})
and the angular resolution is 6\farcs1, 6\farcs4, 6\farcs5 and
12$^{\prime\prime}$ for bands ranging from 3.4 to 22~$\mu$m.

This paper is based on data from the \wise\ All-sky Release, which
comprises images and four-band photometry for over 563 million
sources, and has been publicly available since March 2012. An
object is included in this catalog if it: (1) is detected with
SNR$>$5 in at least one of the four bands; (2) can be measured well
in at least five frames; and (3) is not flagged as a spurious
artifact in at least one band. We refer the reader to the \wise\
All-sky Release Explanatory Supplement for further details\footnote{\wise\
data products and documentation are available at \\ {\tt
http://irsa.ipac.caltech.edu/Missions/wise.html}.} (\citealt{cutri}).

\subsection{Sloan Digital Sky Survey Catalog}
The SDSS (\citealt{york}; \citealt{stoughton}) is a five-band
photometric ($ugriz$ bands) and spectroscopic survey that has mapped
a quarter of the sky, providing photometry, spectra and redshifts
for about a million galaxies and quasars, and photometry for many
more. The imaging reaches 50\% completeness at $r=22.6$
(\citealt{abazajian09}). The SDSS pipeline calculates several kinds
of magnitudes. In this work we have adopted the model magnitudes
(\textit{modelMag}), which perform well for both bright and faint
sources and provide unbiased galaxy colors. Magnitudes are corrected
for Galactic reddening using the dust maps of \citet{schlegel}. When
appropriate, SDSS magnitudes (nearly in the AB system) are converted into 
the Vega system using $m_{\rm AB} = m_{\rm Vega} + t$, where $t$ is estimated by
projecting model stellar spectra into the SDSS $r$-band filter (for details, 
see {\tt kcorrect}\footnote{Available at {\tt
http://howdy.physics.nyu.edu/index.php/Kcorrect}} software). In addition,
SDSS asinh scale magnitudes are converted into Pogson logarithmic scale magnitudes (see 
SDSS website for further details).

\begin{figure}
\epsscale{1.0}
\plotone{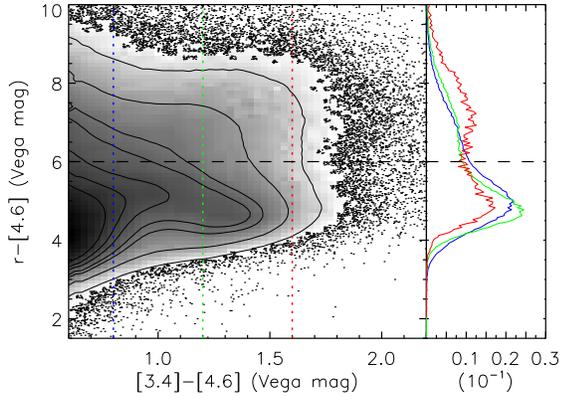}
\caption{\wise-selected AGN split into red (e.g., obscured) sources
with $r - W2 > 6$ and unobscured AGN with $r - W2 \leq 6$. The
grey-scale, contoured region corresponds to high-density regions,
while individual points are shown in areas of low density. Histograms
on the right panel illustrate the marked bi-modality of the distribution
at increasingly redder colors, indicated by the vertical dotted lines
in the left panel.}
\label{fig:rw2_w1w2}
\end{figure}

\begin{figure}
\epsscale{1.0}
\plotone{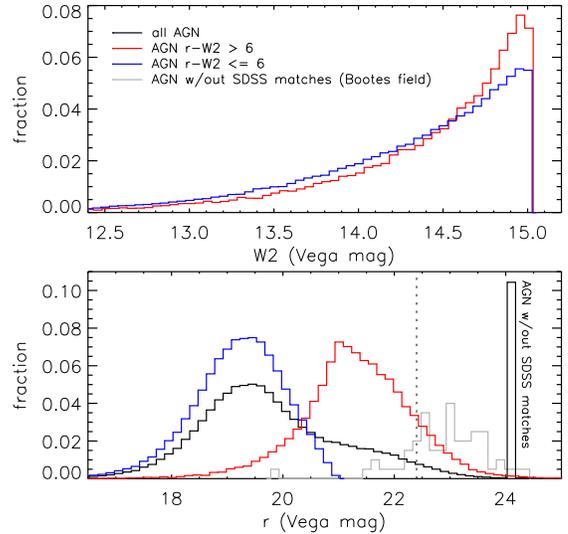}
\caption{{\bf Top:} $W2$ magnitude distribution of \wise\ AGN
candidates split into obscured ($r - W2 > 6$, red line) and unobscured
AGN ($r - W2 \leq 6$, blue line). The former sources are slightly
fainter on average, but both distributions are very similar. {\bf
Bottom:} Optical $r$-band magnitude distribution of \wise\ AGN
candidates showing the effect of the $r - W2$ color cut. For ease
of plotting, the single bin at $r = 24$ represents AGN that lack
an optical match in SDSS; recall that the 50\% completeness limit
of SDSS is at $r = 22.6$ (Vega, Pogson scale; vertical dashed line).
We show the $R$-band magnitudes for 61 such AGN that are in the
deeper Bo\"{o}tes field (grey line); most, in fact, turn out to be
brighter than $r \sim 24$.}
\label{fig:w2magdist}
\end{figure}

\section{WISE-selected AGN}
\subsection{WISE-selected AGN}
Mid-IR selection of AGN relies on distinguishing the characteristic
rising power-law AGN spectrum from the black body spectrum of stellar
populations, which peak at rest-frame 1.6~$\mu$m. This means that
AGN tend to be redder than normal galaxies in the mid-IR.
This was initially shown in {\it Spitzer} data where simple IRAC-band
color cuts isolate AGN from other galaxy populations at $z \simlt
3$ (e.g., \citealt{lacy}; \citealt{stern05}). More recently, the
\wise\ survey has proven very efficient in detecting AGN using just
the two shorter (and more sensitive) bands at 3.4~$\mu$m and
4.6~$\mu$m (\citealt{stern2012}; \citealt{assef2013}). Using
empirical AGN and galaxy spectral templates, \citet{assef} showed
that even pure AGN present typically red $W1 - W2$ colors out to
$z \simlt 3.5$ for reasonable values of dilution by the host galaxy
light (e.g., see Fig.~1 of \citealt{stern2012}). Heavily extincted
AGN are of course even redder in $W1 - W2$. This contrasts with
the bluer $W1 - W2$ colors of: (1) Galactic stars, as only brown
dwarfs with spectral types cooler than T5 have $W1 - W2 > 0.8$
(\citealt{kirkpatrick}); and (2) normal galaxies out to $z \sim
1.2$. Thus, the primary contaminants to the red \wise\ color
selections will be the coolest brown dwarfs, which are quite rare
on the sky, and galaxies at $z \simgt 1.2$, which are effectively
eliminated by our brightness cut, $W2 < 15.05$.

Using \wise\ data over the area covered by the COSMOS survey,
\citet{stern2012} demonstrated that a simple mid-IR color
criterion is extremely robust at selecting AGN candidates. Selecting
sources with $W1 - W2 > 0.8$ above the 10-$\sigma$ flux limit of
0.16~mJy at 4.6~$\mu$m ($W2 < 15.05$, Vega) identifies a large
population of AGN that is $\sim$95\% reliable and nearly 80\%
complete with respect to the {\it Spitzer} AGN selection of
\citet{stern05}. These criteria identify 62 AGN per deg$^2$, as
compared to the $\sim 20$ quasars per deg$^2$ identified by the
optical SDSS quasar selection algorithm, which is sensitive to AGN
of similar intrinsic luminosity (\citealt{richards02}). We
construct our AGN sample by applying the same selection criteria over a
much larger area covered by SDSS. In our sample, we only allow
sources whose $W1$ and $W2$ photometry is unaffected by diffraction
spikes, scattered light, persistence or optical ghosts ($ccflag=0$
in both $W1$ and $W2$). \citet{assef2013} reports on \wise\ selection
of AGN down to $W2 < 17.1$ in the higher ecliptic latitude, and
thus deeper Bo\"{o}tes field. We refer the reader to their work
for a useful comparison of \wise\ AGN selection at various depths. We 
note that, ignoring $W1 - W2$ color for the moment, typical $L^*$ 
galaxies can be observed by \wise\ up to $z \sim 1.2$ at a 5-$\sigma$ 
sensitivity ($W2 = 15.85$; see Fig.~6 of \citealt{yan}). With our 
conservative flux density cut, $W2 = 15.05$, only the brightest, 
several $L^*$ galaxies will be detected by \wise\ at $z \simgt 1$.

Using the selection criteria of $W1-W2>0.8$ and $W2<15.05$, we
selected 176,467 \wise\ AGN candidates over an effective area of
3363~deg$^2$ (see \S\ref{sec:mask} for details about the angular
mask). The $W2<15.05$ magnitude cut guarantees that 99.7\%\ candidates
are detected with SNR$_{\rm W2} > 10$ and that 99.98\% have SNR$_{\rm
W2} > 9$, while the mean SNR$_{\rm W2}$ of the sample is $\sim 20$.

The 176,467 selected AGN candidates are cross-matched with the 
SDSS photometric catalog. Using a matching radius of 1\farcs5, 
we find 152,672 (86.5\%) \wise\ AGN candidates with single optical 
matches, 6095 (3.5\%) sources with two or more SDSS counterparts, 
and 17,700 (10.0\%) \wise\ AGN candidates without an optical source 
listed in the SDSS database. The multiple optical matches are mostly 
due to spurious detections of large sources split into multiple 
components or, in a few cases, real interacting galaxy systems. 
These \wise\ unresolved close galaxy pairs are on scales $\theta<0.001$~deg, 
well below the spatial scales relevant in this work. The clustering analysis 
of galaxies on such small angular scales is beyond 
of the scope of this paper, as it would require full knowledge of the 
deblending performance of the SDSS and \wise\ pipelines. Therefore, we focus 
here on \wise\ AGN candidates with single or no optical counterparts. Note that
so far we have not applied any constraints on SDSS magnitudes, so
that among the 152,672 single \wise-SDSS matches, about 5\% are
fainter than the $r = 22.6$ 50\%-completeness limiting magnitude
of SDSS, but are nevertheless listed in the SDSS catalog. To insure that 
the \wise\ AGN without SDSS counterparts are all real sources and not 
artifacts, we have visually inspected the \wise\ and SDSS images of 
1000 randomly selected objects. We did not find any artifacts from the 
inspection. In addition, Figure~\ref{fig:mask} shows the equatorial 
coordinates of all \wise\ AGN considered in this study, as well as 
the \wise\ AGN lacking optical counterparts in the SDSS database. In this 
latter case, we have closely inspected their spatial distribution projected
on the sky. There are no obvious large scale patterns, suggesting
that the lack of an optical identification is intrinsic to the
sources, and not related to image artifacts, Galactic objects, or
large-scale extinction.

Finally, to further demonstrate that the \wise\ AGN selection is
robustly identifying AGN, we investigate the fraction of \wise-selected
AGN with X-ray counterparts in the 60~ks exposures of the {\it
XMM-Newton} wide-field ($\sim 2$ deg$^2$) survey of the COSMOS field
(XMM-COSMOS -- \citealt{hasinger}; \citealt{brusa}). We find that
$\sim$75\% of \wise-selected AGN are X-ray detected, with the
remaining $\sim$25\% expected to be fainter and/or heavily obscured
AGN missed by the {\it XMM-Newton} observations. Indeed, deeper
{\it Chandra} observations of the central half of the COSMOS field
(\citealt{elvis}) detect 87\% of the \wise-selected AGN. Similar
results were found previously in \citet{stern2012}, though that
work imposed an SNR$_{\rm W2} > 10$ cut, as opposed to the flux
density cut used here.

\begin{figure}
\center
\includegraphics[scale=0.38,angle=90]{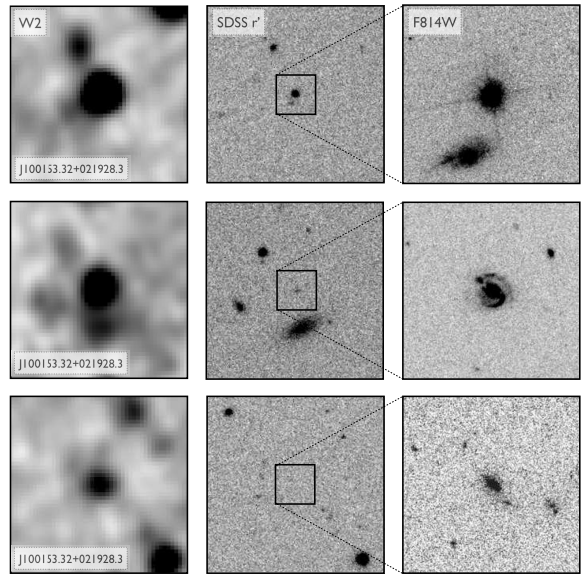}
\caption{Images of three \wise-selected AGN in the COSMOS field,
showing the range of optical morphologies. From left to right, the
columns show \wise\ $W2$ ($\sim 1 \arcmin$ on a side), SDSS $r$
($\sim 1 \arcmin$ on a side), and {\it HST F814W} ($I_{814}$; $\sim
10 \arcsec$ on a side). North is up, and East is to the left. The
top row shows an example of a blue, or unobscured \wise-selected
AGN at $z = 0.372$. The middle row shows an example of an optically
faint, red, or obscured \wise-selected AGN at $z = 0.969$; this
source is X-ray detected and classified as a type-2 AGN
(\citealt{trump}). The bottom row shows an example of the 10\% of
\wise-selected AGN that are undetected by SDSS. This source is
detected by {\it XMM-Newton} and has a photometric redshift of 
$z = 1.512$. See text for further details on the individual sources.}
\label{fig:agn_stamp}
\end{figure}

\begin{figure}
\epsscale{1.0}
\plotone{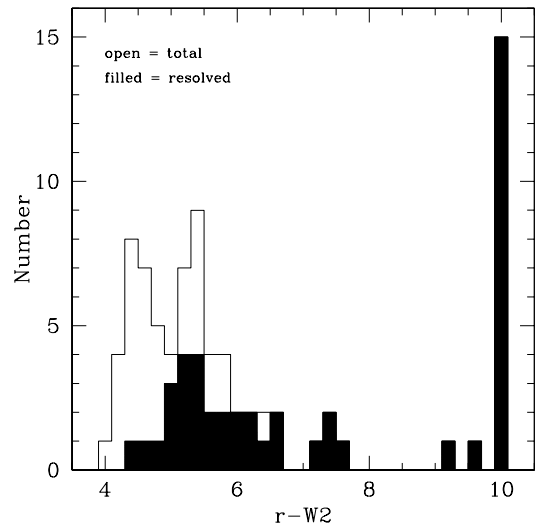}
\caption{Histogram of number of \wise-selected AGN in the {\it
HST}-imaged section of the COSMOS field to our $W2 = 15.05$ depth
as a function of $r -W2$ color. The total (open + filled) histogram
shows the total number of sources, while the filled histogram shows the
subset that are spatially resolved by {\it HST}. Sources that are
undetected by SDSS in the $r$-band are plotted at $r - W2 = 10$.}
\label{fig:morph_hist}
\end{figure}

\begin{figure}
\epsscale{1.0}
\plotone{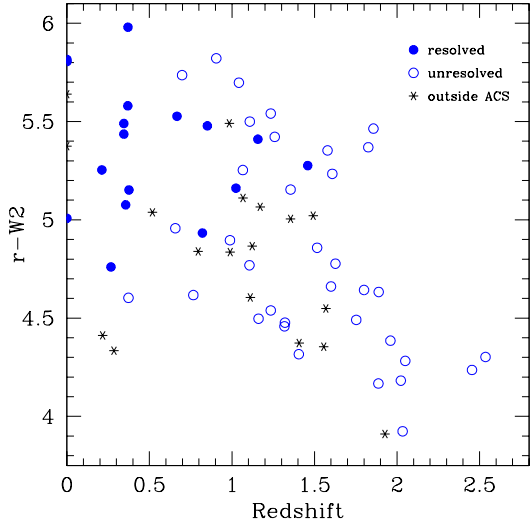}
\caption{$r - W2$ color vs. redshift for blue AGN candidates in the
COSMOS field that have {\it HST} $I_{814}$ morphologies available.
Spatially resolved sources are indicated by filled blue symbols,
while unresolved sources are marked with open symbols. A large
fraction of AGN at low redshift ($z < 0.5$) are clearly resolved
and still meet the blue AGN selection criteria, suggesting that our
low redshift, blue AGN sample is likely a mixture of obscured\textbackslash
unobscured AGN populations.}
\label{fig:morph_z_rw2}
\end{figure}

\subsection{Red and Blue AGN in WISE}
\label{sec:redblue}
As mid-IR observations are relatively insensitive to obscuration
by dust and optical observations are significantly affected by dust
extinction, type-2, or obscured AGN, can be isolated by comparing
\wise\ and SDSS fluxes (\citealt{stern2012}; \citealt{yan}). 
\citet{hickox07,hickox} applied a similar method in the Bo\"{o}tes 
field using IRAC 4.5~$\mu$m and $R$-band photometry to differentiate 
obscured and unobscured AGN. For the sake of completeness, we note, 
however, that there is no rigorous and unique definition to 
differentiate obscured and unobscured AGN across all wavelengths.

In this work, we divide the \wise\ AGN sample according to $r - W2$
color. Figure~\ref{fig:rw2_w1w2} illustrates that AGN show a bimodal
color distribution that separates two populations of AGN. Those
with colors redder than $r - W2 = 6$ are, of course, optically
faint (or undetected in SDSS), but nevertheless well detected at
4.6~$\mu$m. We call these ``red AGN'', in contrast with ``blue
AGN'' that are relatively bright at both mid-IR and optical
wavelengths. We fold AGN lacking optical matches into the
red AGN sample. As shown in \citet{hickox07}, the red population is 
more closely associated with type-2 AGN, while the blue population is 
associated with type-1 AGN, e.g., AGN presenting broad emission lines 
in optical spectroscopy. In total, about 60,000 sources are selected 
as red AGN candidates,
implying a type-1 fraction of roughly 55\%, similar to the fraction
found by \citet{assef2013} for luminous AGN with bolometric
luminosities exceeding a few times 10$^{46}$~erg~s$^{-1}$. In
\S\ref{sec:zdist_sel} we evaluate the model selection function of
red and blue AGN to test the reliability of the $r - W2$ criteria
to separate type-1 and type-2 AGN.

Figure~\ref{fig:w2magdist} (top panel) shows the $W2$ magnitude
distribution of red and blue \wise-selected AGN. Although red AGN
seem slightly fainter at mid-IR wavelengths in general, both
subsamples have similar distributions, suggesting there is no strong
bias due to the $r - W2$ color cut. The bottom panel shows the
distribution of SDSS $r$-band magnitudes (corrected for Galactic 
reddening, converted to Vega and in the Pogson scale). Blue AGN are 
considerable brighter, peaking at $r \sim 19.3$ and falling steeply 
at $r \simgt 19.5$. Most red AGN are fainter, peaking around 
$r \sim 21.2$ and extending to fainter magnitudes, beyond the nominal 
SDSS completeness limit. A considerable fraction (10\%) of \wise\ AGN 
candidates are simply
undetected by SDSS; we indicate such sources with a single bin at
$r = 24$. The Bo\"{o}tes field has considerably deeper $R$-band
photometry available, reaching $R \sim 26$ (5-$\sigma$, point source;
\citealt{jannuzi}). There are 61 SDSS-undetected, \wise-selected
AGN in Bo\"{o}tes. Their $R$-band magnitude distribution peaks at
$R \sim 23$ (gray line) with all sources having optical counterparts.
This again illustrates the optical faintness, but detectability of
essentially all \wise-selected AGN.
Finally, we note that the fraction of \wise-selected AGN with X-ray 
counterparts in the {\it XMM-Newton} wide-field at the 0.5$-$10~keV band
is $\sim 83\%$ for blue AGN ($r - W2 \leq 6$), dropping to $\sim 68\%$ for 
red AGN ($r - W2 > 6$). These high detection rates further demonstrate 
the reliability of our sample.

\begin{figure}
\epsscale{1.0}
\plotone{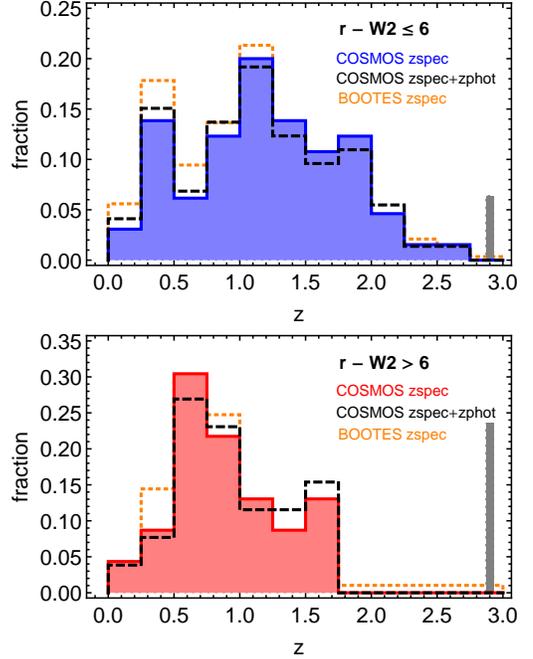}  
\caption{Redshift distribution of \wise-selected AGN in COSMOS. The
top panel highlights the blue AGN ($r - W2 \leq 6$), with the solid
histogram showing sources with spectroscopic redshifts and the
dashed histogram including five additional photometric redshifts.
The five sources lacking both spectroscopic and photometric redshifts
are plotted at $z = 2.9$ (gray bar).  The bottom panel highlights
the red AGN, again distinguishing spectroscopic redshifts (solid
histogram) and photometric redshifts (dashed). The eight sources
lacking both spectroscopic and photometric redshifts are plotted
at $z = 2.9$ (gray bar). For reference, we also show in both panels
the corresponding redshift distributions in Bo\"{o}tes (dotted
orange).}
\label{fig:zhist}
\end{figure}

\begin{figure}
\epsscale{1.1}
\plotone{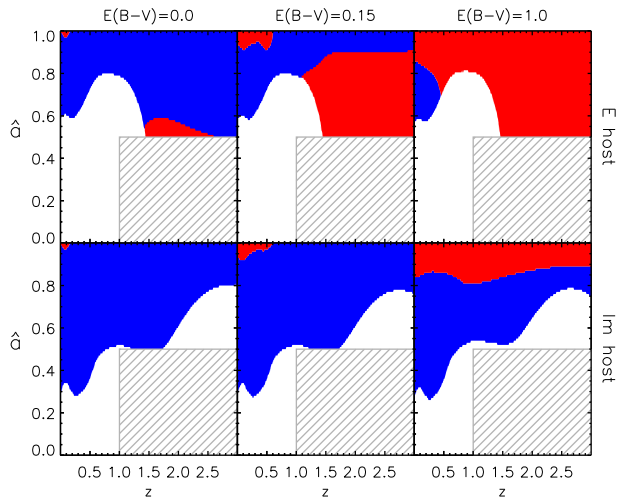}
\caption{Model selection function of blue and red AGN constructed
using mock objects that adopt the AGN and galaxy SED templates of
\citet{assef}. The parameter $\hat{a}$ is the fraction of the
bolometric luminosity coming from the AGN component (see text for
details). Each panel shows for a given host galaxy type (E or Im)
and reddening value, whether an object would be targeted as a blue
AGN (blue region), red AGN (red region), or an inactive galaxy
(white region). The gray hatched area marks the region where \wise\
is not sensitive due to its shallowness given our $W2 < 15.05$
brightness cut, namely $z > 1$ host-galaxy dominated objects.  While
essentially all of the unobscured AGN (left panels) are correctly
identified as blue AGN, some fraction of obscured AGN (right panels)
will have blue AGN colors.  Phrased differently, we expect the red
AGN sample to be a relatively pure sample of obscured AGN, while
the blue AGN sample will primarily be unobscured AGN, but will have
some contamination from obscured sources.}
\label{fig:ahat}
\end{figure}

\begin{figure}
\epsscale{1.0}
\plotone{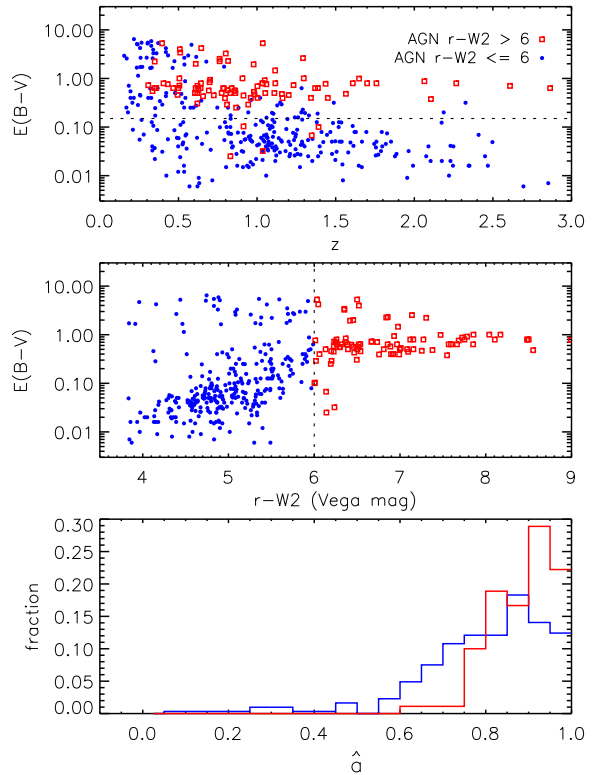}
\caption{Dependence of the reddening parameter, $E(B-V)$, with
redshift (top panel) and $r - W2$ color (middle panel) for blue (filled 
symbols) and red (open squares) \wise-selected AGN candidates in the 
Bo\"{o}tes field. $E(B-V)$ is derived by fitting the AGN and galaxy 
SED templates of \citet{assef}. In general, red AGN tend to have 
considerable reddening at all redshifts while blue AGN are mostly 
unreddened at $z > 0.5$, but can have large reddening values at 
lower redshift. The fiducial type-1/type-2 separation is around 
$E(B-V) = 0.15$. The bottom panel shows the distribution of the
\^{a} parameter (see \S\ref{sec:zdist_sel} for definition) for
red and blue AGN.}
\label{fig:reddz}
\end{figure}

\subsection{Morphologies}
\label{sec:morph}
Figure~\ref{fig:agn_stamp} shows the range of optical morphologies
of \wise-selected AGN. For three candidates in the COSMOS field,
we display $\sim 1 \arcmin$ on a side images in \wise\ $W2$ and
SDSS $r$-band, and $\sim 10 \arcsec$ on a side {\it Hubble Space
Telescope} ({\it HST}) images in the $F814W$ filter. The top row
shows an example of a blue, or unobscured \wise-selected AGN:
WISE~J100025.25+015852.1 is an optically bright, optically unresolved
SDSS quasar ($r-W2=4.6$) at redshift $z = 0.372$. The
middle row shows an example of an optically faint, or red, obscured
\wise-selected AGN: WISE J100005.98+015453.1 is an optically faint
source detected by SDSS ($r - W2 = 6.7$). \citet{trump} report a
redshift of $z = 0.969$ for this X-ray detected, optically resolved
source and classify it as type-2 AGN based on its spectrum. The
bottom row shows an example of the 10\% of \wise-selected AGN which
are undetected by SDSS: WISE J100153.32+021928.3 is undetected by
SDSS ($r - W2 \geq 7.5$), but is detected by both {\it HST} and 
{\it XMM-Newton}. The source has a photometric redshift 
of $z = 1.512$.

Optical morphologies offer an additional observable with which to
investigate the \wise\ AGN selection. Luminous, unobscured, or
type-1 AGN are typically unresolved at optical wavelengths, which
was one of their foundational attributes that led to the name
``quasar'', or quasi-stellar radio source. We have known for several
decades now that only $\sim 15\%$ of quasars are radio-loud, with
little variation in this fraction with either redshift or optical
luminosity, at least at the high luminosity end (e.g.,
\citealt{stern2000}). Similarly, mid-IR selection is showing that
unresolved, unobscured quasars represent a minority population of
luminous AGN. Indeed, using the SDSS {\tt type} flag to discriminate
morphologies, we find only $\sim55\%$ of the \wise-selected AGN
considered in this paper are classified as unresolved point sources.

We use the COSMOS field to characterize how morphology depends on
optical-to-mid-IR color for \wise-selected AGN.
Figure~\ref{fig:morph_hist} shows a histogram of the optical-to-mid-IR
colors of the 82 \wise-selected AGN with {\it HST} ACS (F814W) imaging in COSMOS
to our $W2 = 15.05$ depth, coded by optical morphology. Fifteen
of the sources are undetected by SDSS in the $r$-band, and are
simply plotted at $r - W2 = 10$; all 15 of these sources are detected
in the deeper {\it HST} $F814W$ imaging and are spatially resolved.
Indeed, of the 28 red AGN candidates with $r - W2 > 6$, only 1 (4\%)
is unresolved. This supports our expectation that red optical-to-mid-IR
colors select a clean sample of obscured AGN with little contamination
from unobscured AGN.

Of the 54 blue AGN candidates, 35 (65\%) are unresolved.
Figure~\ref{fig:morph_z_rw2} shows the $r - W2$ color vs. redshift
for these blue AGN candidates, with symbols indicating their {\it
HST} morphologies. As we will show in the next section, most of
the resolved AGN -- e.g., likely obscured AGN contaminating our
blue AGN selection -- are at lower redshift ($z < 0.5$) and, in
fact, reside in the redder end of our blue AGN selection. However,
Figure~\ref{fig:morph_hist} also clearly shows that it is not
feasible to simply make a bluer $r - W2$ cut to separate obscured
(e.g., resolved) and unobscured (e.g., unresolved) AGN.

To characterize the host galaxies of \wise-selected candidates we
performed more detailed visual classifications on the {\it HST} ACS
image cutouts (independently by three of the four authors; as we
agreed for the majority of objects, we report the average here).
For red AGN, we find that 54\% (15) are disk galaxies or interacting
systems, 32\% (9) are elliptical or point sources, and the remaining
14\% (4) have uncertain morphology. This contrasts with blue AGN,
where we find that 20\% (11) are disk galaxies and 80\% (43) are
either point sources (most) or ellipticals. These results are
consistent with the work of \citet{griffith}, who studied the
morphology of AGN in COSMOS selected at radio, X-ray and mid-IR
wavelengths.  That work found that the red mid-IR-selected AGN
consist of 63\% disk galaxiess, 22\% point sources/ellipticals, and
15\% other morphology, while the blue AGN consist of 15\% disk
galaxies and 85\% point sources/ellipticals. The main conclusion
we wish to draw here is that given its high fraction of disk galaxies,
the red AGN sample is not dominated by typical red sequence galaxies.
In fact, the red AGN have a higher fraction of disk galaxies than
the blue AGN.  As further discussed in \S\ref{sec:hostgal}, this
suggests it is unlikely that a bias in host galaxy type (favoring
red AGN in early-type hosts and blue AGN in late-type hosts) could
have a large impact in the interpretation of the clustering results
presented in \S\ref{sec:clust_results}.

\subsection{Redshift Distribution and Selection Function}
\label{sec:zdist_sel}
Given the difference in optical flux introduced by the $r - W2$
cut, it is not unreasonable to expected differences in the redshift
distribution of blue and red AGN. In order to understand the
redshift distribution and properties of \wise\ AGN candidates, we
have matched our list to publicly available spectroscopy in the
COSMOS field as well as recent spectroscopic observations (see
\citealt{stern2012} for details about the compiled list of spectroscopic
and photometric redshifts).

\begin{figure}
\epsscale{1.0}
\plotone{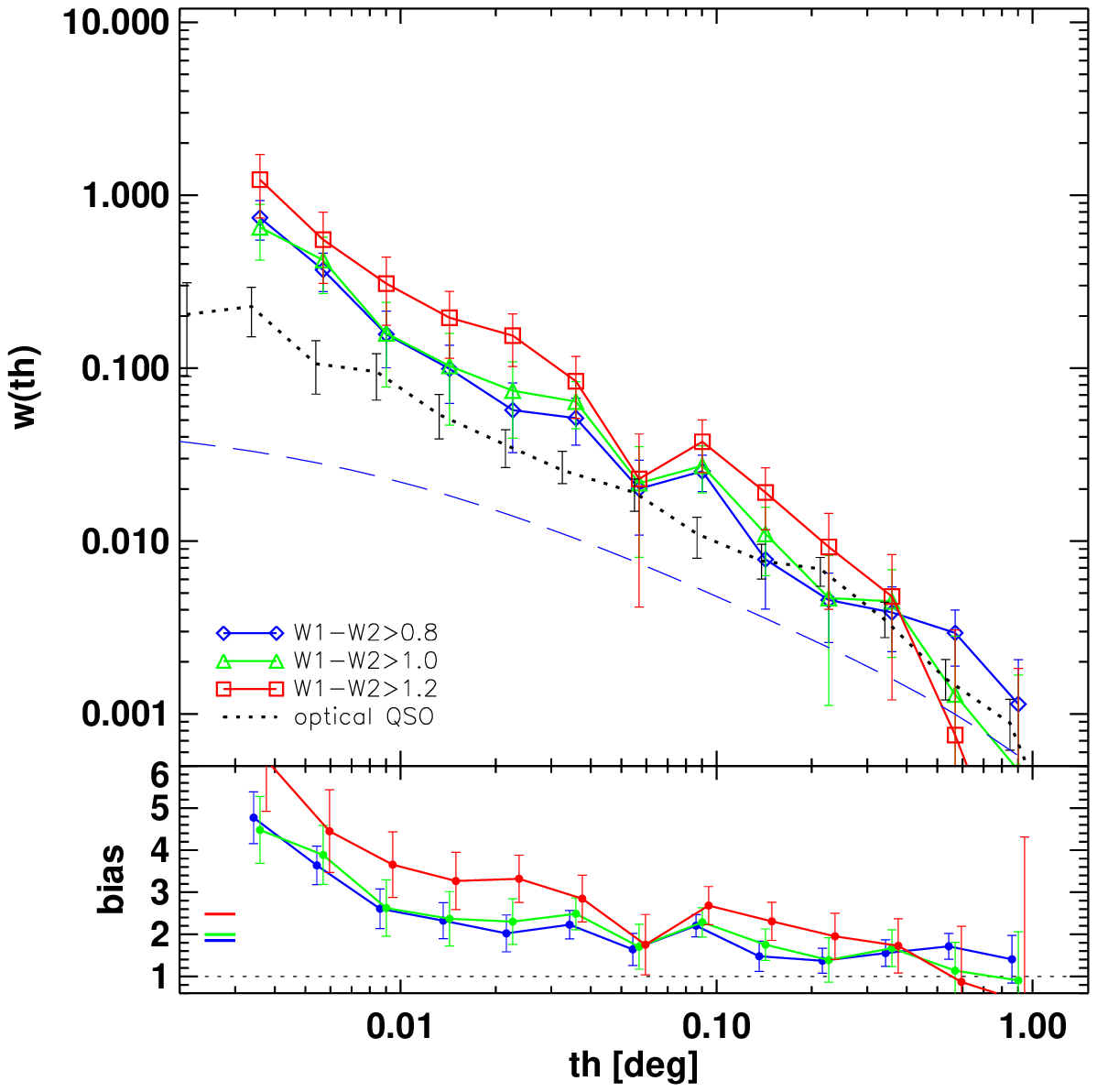} 
\caption{Angular correlation function $w(\theta)$ of \wise-selected
AGN with increasingly red $W1 - W2$ color cuts. For reference, we
also show data for optical quasars at $z_{\rm phot} < 2.3$ from 
\citet[][dotted line]{myers07}. Model predictions for the dark matter angular 
correlation function, $w_{\rm dm}(\theta)$  (dashed line) are computed 
using the \citet{peacock96} fitting function and the same AGN redshift 
distribution as the $W1 - W2 > 0.8$ sample. The bottom panel shows the
absolute bias $b=\sqrt{w/w_{\rm dm}}$. Markers on the left indicate
the mean bias value over the range 0.02-0.4~deg.}
\label{fig:acf_w1w2}
\end{figure}

Figure~\ref{fig:zhist} shows the redshift distribution of the 112
\wise-selected AGN available in COSMOS, of which 88 have spectroscopic
redshifts and 11 have photometric redshifts (plus 13 objects with no
redshift information available). The top panel highlights
the blue AGN ($r - W2 \leq 6$), including five sources, plotted at
$z=2.9$, that lack both spectroscopic and photometric redshifts. The
distribution peaks around $z \sim 1.1$ and extends up to $z \sim
2.5$, with most of the sources at $0.8 < z < 2$. There is an
indication of a second smaller peak at $z \sim 0.5$, most probably
(as we will see later) due to type-2 AGN that enter into the redder
part of our blue sample selection at low redshift. For reference,
we also show the spectroscopic redshifts of 536 \wise\ AGN candidates
within the Bo\"{o}tes field (dashed histogram), obtained from the
AGN and Galaxy Evolution Survey (AGES; \citealt{kochanek}). This
survey has different completeness levels for different galaxy samples
($I < 20$ for galaxies, $I < 22.5$ for AGN, but with varying priority
levels based on their brightness at mid-IR, $24 \mu$m, radio, and
X-ray energies) and therefore a complicated redshift selection
function. However, considering the differences in target selection
as compared to COSMOS (which essentially targeted every source to
$R \sim 25$), the two distributions agree remarkably well. This
suggests that both are not far from the true redshift distribution
of \wise-selected AGN with blue $r - W2$ colors. The bottom panel
in Figure~\ref{fig:zhist} shows the corresponding distributions for
\wise-selected AGN with red $r - W2$ colors, including eight sources
lacking both spectroscopic and photometric redshifts plotted at $z = 2.9$. 
Red AGN candidates peak at lower redshift, around $z \sim 0.8$, and 
extend up to $z \sim 1.8$. Again, the agreement with AGES redshifts 
in the Bo\"{o}tes field is notable.

To further understand the nature of the differences in redshift
among the red and blue AGN samples, we model their selection function
by constructing mock objects using the AGN and galaxy spectral
energy distribution (SED) templates from \citet{assef}. The parameter
$\hat{a} \equiv L_{\rm AGN}/(L_{\rm host}+L_{\rm AGN})$ quantifies
the fraction of the bolometric luminosity coming from the AGN
component (see \citealt{assef}, 2013 for details). Figure~\ref{fig:ahat}
shows whether an object with a given host galaxy type (E or Im),
$\hat{a}$ value, and reddening towards the accretion disk, parametrized
by $E(B-V)$, would be targeted as a blue AGN ($r - W2 \leq 6$, blue
region), as a red AGN ($r - W2 > 6$, red region), or as an inactive
galaxy ($W1 - W2 < 0.8$, white region). As expected, at low
$\hat{a}$, most systems are characterized as normal galaxies. The
panels at $E(B-V)=0.0$ and $E(B-V)=1.0$ highlight the extreme cases
of a zero reddening or a heavily extincted AGN; the typical
boundary between type-1 and type-2 AGN corresponds to a reddening
of $E(B-V)=0.15$ (see \citealt{assef2013} for details). The gray
hatched area marks the region where \wise\ is not sensitive due to
the shallowness imposed by our $W2 < 15.05$ flux density requirement,
namely $z \simgt 1$ host-galaxy dominated objects. This figure shows
that while it is very unlikely to misclassify a blue AGN as a red
one, the opposite happens for a significant fraction of parameter
space, suggesting that our red AGN selection constitutes a reliable
yet incomplete type-2 AGN sample, while our blue sample consists
of a mixture of type-1 and type-2 AGN.

We also used the deep, multi-wavelength data available in the
Bo\"{o}tes field to do detailed SED modeling of \wise-selected AGN
and explore how reddening relates to $r - W2$ color for blue and
red AGN candidates as a function of redshift. This is shown in
Figure~\ref{fig:reddz}. The reddening parameter $E(B-V)$ is derived 
by fitting the AGN and galaxy SED templates of \citet{assef}. As 
expected, red AGN tend to show considerable reddening at all redshifts, 
with $E(B-V) \simgt 0.7$, while blue AGN are mostly unreddened above 
$z \sim 0.5$. However, below this redshift, blue AGN can sometimes 
show large reddening values, consistent with the idea that some of 
these objects might well be type-2 AGN interlopers in the blue sample.
As shown in the bottom panel of Figure~\ref{fig:reddz}, the distributions
of $\hat{a}$ are strongly peaked toward high values, with most red AGN
above $\hat{a}\sim0.8$, and a minor fraction of blue AGN 
with $0.6<\hat{a}<0.8$. This means that while the blue area in the bottom 
right panel of Figure~\ref{fig:ahat} is large, only a minority of sources
could be potentially biased due to the galaxy host type (i.e. selected as 
blue AGN due to the presence of Im galaxy host).

\section{Angular Correlation Analysis}
\subsection{The Angular Correlation Function}
\label{sec:mask}
A standard tool to measure galaxy clustering is the two-point angular
correlation function, $w(\theta)$. It is defined as the probability
that a given pair of galaxies separated by an angle $\theta$ on the
sky are contained within a solid angle $d\omega$
\begin{equation}
 dP=n(1+w(\theta))d\omega,
\end{equation}
where $n$ is the mean number density of galaxies. In practice,
$w(\theta)$ is calculated by counting pairs of galaxies in annuli
of different radii and comparing with the corresponding counts in
a random sample of galaxies. To estimate $w(\theta)$ we use the
\citet{landy93} estimator, given by
\begin{equation}
 w(\theta)=\frac{DD-2DR+RR}{RR},
\end{equation}
where $DD$, $DR$ and $RR$ are the normalized data-data, data-random
and random-random pair counts, respectively. It is very important
that the random sample has the same angular selection as the data
pairs. For this purpose we constructed an angular mask using the
software {\tt mangle}\footnote{Available at {\tt
http://space.mit.edu/$\sim$molly/mangle/}.} that describes
the survey geometry in terms of disjoint spherical polygons. This
mask accounts for the holes caused by bad quality fields in the
SDSS survey, as well as the areas around bright stars selected from
the Tycho 2 catalog ($BTMAG < 11.5$). In addition, we also remove
the areas around large ($>2\arcsec$) sources from the 2MASS Extended
Source Catalog that in some cases appear decomposed into multiple
sources in \wise. Galactic absorption can have an impact in
faint galaxy counts (\citealt{myers06}), so we mask out areas with
$A_g > 0.18$. Finally, we avoid regions contaminated by the Moon
and limit the sample to the rectangular area bounded by 
$135^\circ <$ R.A. $<226^\circ$ and $1^\circ <$ Dec. $< 54^\circ$ (J2000). 
These rather conservative limits avoid both the Galactic plane, where 
contamination by stars could present an issue, and the ecliptic pole, 
where the sensitivity of \wise\ improves substantially due to denser 
coverage and lower zodiacal background. Our selected area has a typical 
\wise\ coverage of $\sim 13$ frames per bandpass.

\subsection{Absolute Bias and Halo Masses of WISE AGN}
\label{sec:halomass}
At small scales, the clustering of an extragalactic source population
is difficult to predict due to processes such as merging and
interactions. However, at larger scales (e.g., $>$ 1-2 $h^{-1}$~Mpc),
galaxy interactions have little impact and the galaxy correlation
function follows that of the dark matter halos.  At any redshift,
massive halos cluster more strongly than less massive halos. Given
an average redshift, this, in turn, allows one to estimate the
typical mass of dark matter halos in which objects reside by
estimating their absolute bias, i.e., their observed clustering
level with respect to that of the underlying dark matter.

We compare our $w(\theta)$ measurements to the predictions of the
standard cold dark matter (CDM) model in the linear perturbation
theory of structure growth along with the non-linear correction.
To compute the dark matter angular two-point correlation function,
$w_{\rm dm}(\theta)$, we use the non-linear fitting function of
\citet{peacock96} for the CDM power spectrum projected onto the
same AGN redshift distribution. The bias factor is simply defined
as $b \equiv (w(\theta)/w_{\rm dm}(\theta))^{1/2}$. In general, the
bias is a function of scale, but under the assumption that galaxies
cluster in a similar manner as dark matter, the bias factor is
nearly scale-independent. This is particularly valid in the linear
regime (i.e., large scales; see \citealt{verde}). We limit the bias
and the corresponding fits from $\theta=0.04^\circ$ to $\theta=0.4^\circ$,
which corresponds to scales of roughly $\sim 800\, h^{-1}$~kpc
to $\sim 8\, h^{-1}$~Mpc at $z \sim 1.2$.

\begin{figure}
\epsscale{1.0}
\plotone{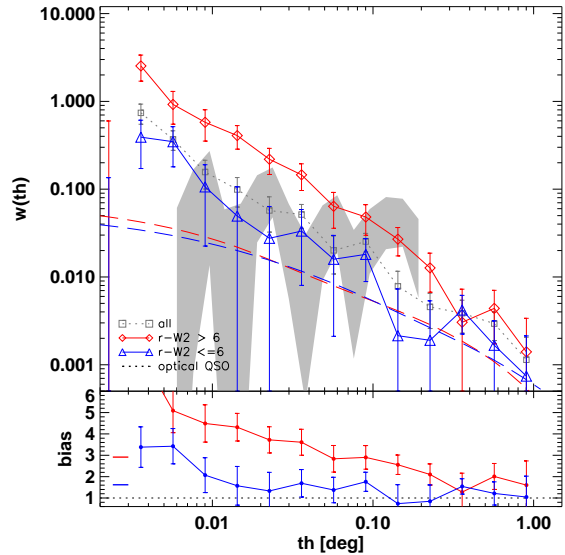} 
\caption{Angular correlation function $w(\theta)$ of \wise-selected
AGN split into obscured sources with $r - W2 > 6$ and unobscured
AGN with $r - W2 \leq 6$. The bottom panel shows the absolute bias with
respect to the dark matter angular correlation (dashed line). Markers
on the left indicate the mean bias value.  The grey shaded region
shows the angular autocorrelation of type-1 quasars from \citet{hickox}
(inferred from the quasar-galaxy and galaxy-galaxy correlation 
function), which is in broad agreement with our estimation for the 
blue AGN sample.}
\label{fig:acf_rw2}
\end{figure}

\begin{figure}
\epsscale{1.0}
\plotone{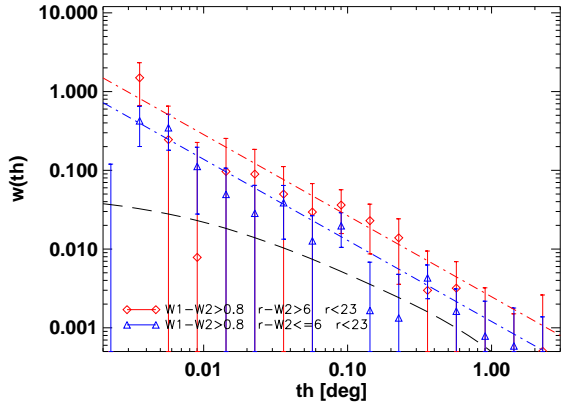}
\caption{Angular correlation function $w(\theta)$ of \wise-selected
AGN as in Figure~\ref{fig:acf_rw2}, but limited to sources with
$r$-band counterparts brighter than $r = 23$ in the SDSS survey.
Simple power-law fits of the form $A\theta^{-\gamma}$ (dot-dashed
lines) have a correlation amplitude a factor of $\sim 2$ larger for the
obscured population compared to the unobscured sources.}
\label{fig:acf_rw2_r23}
\end{figure}

Using an ellipsoidal collapse model, \citet{sheth} related the halo
bias factor to its mass and calibrated a fitting relation for a
large library of cosmological $N$-body simulations: 
\begin{equation}
\begin{multlined}
b(M_{\rm halo},z) = 1 + \frac{1}{\sqrt a \delta_{c}(z)} [ \sqrt a (a\nu^2) \\
\shoveleft[1cm]{+ \sqrt a b (a\nu^2)^{1-c}} - \frac{(a\nu^2)^c}{(a\nu^2)^c + b(1-c)(1-c/2)}] \\
\end{multlined}
\end{equation}
where $a=0.707$, $b=0.5$, $c=0.6$ and $\delta_c(z)$ is the critical
density ratio for collapse given by \citet{navarro} as
$\delta_c(z)=0.15(12\pi)^{2/3}\Omega_{mz}$, and $\Omega_{mz} \equiv
(H_0/H(z))^2 \Omega_m (1+z)^3$. $H(z)$ depends on the cosmology as
$H^2(z)=H^2_0 (\Omega_m(1+z)^3 + \Omega_\Lambda)$, and $\nu$ is
defined as $\nu \equiv \delta_c{z}/\sigma(M)D(z)$, where $D(z)$ is
the linear growth factor, here approximated analytically using the
formulae by \citet{carroll}. The rms fluctuation of the density
field is given by
\begin{equation}
\sigma^2(M_h)=\frac{1}{2\pi^2}\int_{0}^{\infty}k^2P(k)\left [ \frac{3(\textrm{sin}(kr)-(kr) \textrm{cos}(kr))}{(kr)^3}\right]dk,
\end{equation}
where the term in brackets represents the spherical top-hat window
function (\citealt{peebles}) and the radius $r$ is related to the
enclosed halo mass $M_h$ as
\begin{equation}	
r=\sqrt[3]{\frac{3M_h}{4\pi\rho_0}},
\end{equation}
where $\rho_0$ is the present mean density of the Universe, given
by $2.78\times 10^{11}\, \Omega_m\, h^2\, M_\odot\, \textrm{Mpc}^{-3}$.
The linear power spectrum of density fluctuations, $P(k)
\propto T^2(q)\, k^n$ with $n$=1 (the Harrison-Zel'dovich case), is
constructed using the fitting formula of \citet{eisenstein} for the
transfer function $T(k)$ and normalized with the adopted value of
$\sigma_8 =0.84$ for $r = 8\, h^{-1}$ Mpc.

\section{Results}
\label{sec:clust_results}
\subsection{Comparison to Optically Selected Quasars}
We begin our analysis by exploring the angular clustering for the
full sample of AGN selected by \wise. Figure~\ref{fig:acf_w1w2}
shows that AGN with $W1 - W2 > 0.8$ present an angular correlation
similar to that of optical quasars selected from SDSS by \citet{myers07}
using a photometric kernel density estimation (KDE) technique
(\citealt{richards04}). A power-law fit of the form $w(\theta) = A
\theta^{-\gamma}$ gives a value of $\gamma = 1.03 \pm 0.11$ within
the range $\theta= [0.02^\circ - 0.5^\circ]$ ($\sim 0.4-10\, h^{-1}$~Mpc at
$z = 1.1$). \citet{myers06} find $\gamma= 0.98 \pm 0.15$ for
optically selected quasars at $z = 1.4$, while \citet{croom05} find
a slightly shallower value for 2QZ quasars, $\gamma= 0.86 \pm 0.06$
when averaged over scales of 1-100~$h^{-1}$~Mpc and after correcting
for redshift distortions. These slight differences are not entirely
surprising considering the very different AGN selection criteria
and the fact that the \wise\ AGN sample includes both obscured and
unobscured AGN, while the optical quasar samples are entirely comprised 
of broad-lined, type-1 AGN. Furthermore, the clustering of quasars might 
not be properly represented by a single power law. 

At scales below $\theta \sim 0.1^\circ$, we find that redder AGN
have slightly higher angular clustering. This is interesting
considering that this scale ($\sim 2\, h^{-1}$~Mpc) marks the
transition between the 1-halo and 2-halo terms, which, in the
framework of halo clustering models, arises from galaxy pairs located
in either the same or in two different halos, respectively. As shown
at the bottom panel of Figure~\ref{fig:acf_w1w2}, the absolute bias
for \wise-selected AGN with $W1 - W2 > 0.8$ with respect to the
underlying dark matter distribution is $b = 1.9 \pm 0.4$, as compared
to $b = 2.5 \pm 0.6$ for \wise-selected AGN with redder mid-IR
colors, $W1 - W2 > 1.2$. Taking into account the caveat that different
redshift and luminosity distributions can possibly bias the results,
the simplest interpretation is that redder AGN are hosted by slightly
more massive dark matter halos. For type-1 AGN at $z\lesssim2.5$
previous work has shown that the clustering depends only weakly on
redshift, luminosity or color (\citealt{shen09}; \citealt{ross09}).
However, for type-2 AGN this is mostly unknown and our $W1 - W2 >
0.8$ sample is expected to be a mixture of both type-1 and type-2
AGN.  Finally, we note that our results compare well to the bias
estimates obtained by \citet{myers07} for optical quasars over a
similarly broad redshift range centered at $\langle z \rangle=1.4$.

\subsection{Clustering of Red and Blue AGN}
We explore now the angular clustering of \wise-selected red and
blue AGN.  The corresponding correlation functions, shown in
Figure~\ref{fig:acf_rw2}, display very different amplitudes. For a
fixed slope $\gamma = 1.03$ (that of the entire AGN sample), blue,
or unobscured AGN (e.g., $r - W2 \leq 6$) have $A = 0.0010 \pm
0.0002$, while red, or obscured AGN (e.g., $r - W2 > 6$) have $A =
0.0039 \pm 0.0004$, i.e., a factor of $\sim4$ larger. The bottom
panel shows that the mean bias of obscured sources relative to the
dark matter is $b = 2.9 \pm 0.6$, as compared to $b = 1.6 \pm 0.6$
for unobscured AGN. For reference, we also show in Figure~\ref{fig:acf_rw2}
the angular clustering of type-1 quasars (grey shaded area) from
\citet{hickox}, which is in broad agreement with our estimation for
the blue AGN sample.

Part of the difference in clustering strength could, in principle,
be due to the obscured sources having a different selection, that
is, since obscured sources are required to be optically faint (or
undetected), they could reside at slightly higher redshifts than
their unobscured cousins. On the contrary, spectroscopy from both 
COSMOS and  Bo\"{o}tes demonstrates that red AGN tend to be at slightly
lower redshift (Figure~\ref{fig:zhist}).  The caveat is that there
is a $\sim20\%$ incompleteness in the two spectroscopic samples and
the sample sizes are not extremely large.  While directly
comparing the full and complete redshift distributions for blue and
red \wise-selected AGN would be ideal to check whether their different
clustering strengths are related to different redshift distributions,
we can nevertheless minimize it by selecting AGN limited only to
those with $r < 23$ counterparts in SDSS. The amplitudes of the
best-fit power-law become $A = 0.0024 \pm 0.0006$ for obscured AGN,
compared to $A = 0.0012 \pm 0.0002$ for unobscured AGN, for a fixed
slope $\gamma = 1.03$. Figure~\ref{fig:acf_rw2_r23} shows the
corresponding angular auto-correlations, illustrating once again
that, while noisier, obscured AGN have a correlation amplitude a
factor of $\sim 2$ larger than the unobscured sources.

\begin{figure}
\plotone{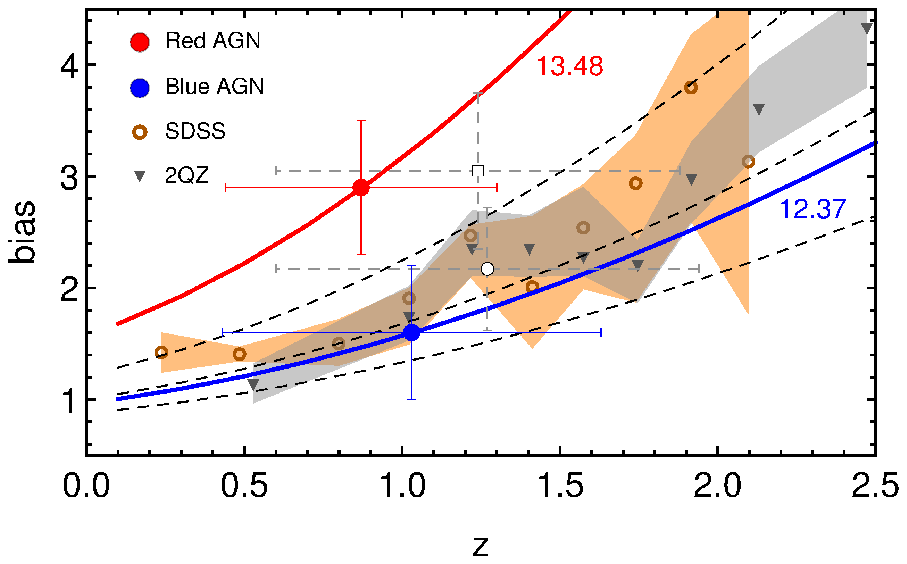}
\caption{Bias as a function of redshift for \wise\ blue and red AGN, shown 
at the mean redshift of their corresponding best-fit distributions. For 
reference, we also overlay data derived from optical SDSS quasars (orange, 
\citealt{ross09}) and 2QZ quasars (gray, \citealt{croom05}), as well as 
previous results from \citet{hickox} for obscured (hollow square) and 
unobscured AGN (hollow circle). Dashed lines are models of constant halo 
mass of $\log M/M_\odot\, h^{-1} = 13, 12.5, 12$ (from top to bottom), 
while the best-fit cases for \wise\ AGN are indicated by solid, thick lines.}
\label{fig:biasz}
\end{figure}

Given the difference in amplitude between the correlation functions
of red and blue AGN, we investigate how this reflects into the
masses of dark matter halos that host them. Using the prescriptions
described in \S\ref{sec:halomass}, we estimate that blue AGN at
$z\sim 1$ are hosted in halos of characteristic mass $\log(M/M_\odot\,
h^{-1})=12.37_{-1.00}^{+0.57}$.  This is in excellent agreement
with the halo mass of $\log(M/M_\odot\, h^{-1}) \sim 12.3$ reported
by \citet{ross09} for SDSS optical quasars at $z<2.2$. \citet{croom05}
finds a similar value of $\log(M/M_\odot\, h^{-1}) \sim 12.5_{-0.3}^{+0.2}$
for 2QZ quasars hosts. In Figure \ref{fig:biasz} we show the bias
as function of redshift for the best-fit model (thick blue line),
along with models of constant halo mass (dashed black lines) for
reference. We find that the halos of our red AGN have a much larger
characteristic mass of $\log(M/M_\odot\, h^{-1}) = 13.48_{-0.31}^{+0.54}$,
i.e. over a factor of 10 larger than for blue AGN. We discuss
the physical implications of this result in the following section.
We also note that \citet{hickox} reports a very similar mass of
$\log(M/M_\odot\, h^{-1}) = 13.3_{-0.4}^{+0.3}$ for their obscured
quasar sample, though their value of $\log(M/M_\odot\, h^{-1}) =
12.7_{-0.6}^{+0.4}$ for unobscured quasars is slightly larger than
both our value and literature results for optically selected
unobscured quasars.

\subsection{The Host Galaxies of WISE AGN}
\label{sec:hostgal}
To understand the clustering result of our red and blue samples,
we study the host galaxies of \wise-selected AGN using SED fitting
and the morphology classifications discussed in \S\ref{sec:morph}.
This is important because the observed difference in clustering
might, in principle, be attributed to a selection effect that biases
our red AGN sample to being hosted by early-type galaxies and our
blue AGN to being hosted by late-type galaxies. Such a difference
might either be the result of an intrinsic difference between the
populations or due to a selection function bias. In particular,
Figure \ref{fig:ahat} suggests that our red AGN sample could be
biased against type-2 quasars in starburst galaxies if mid-IR
selected AGN had a large spread over \^{a} values.

First, we use the SED fitting of \wise-selected AGN candidates in
the Bo\"{o}tes field with the templates of \citet{assef} to analyze
the distribution of host light coming from each of the three galaxy
templates (E, Sbc and Im). From Figure \ref{fig:reddz}, the blue
AGN sample contains some sources with considerable dust obscuration
(i.e., well above the $E(B-V)=0.15$ boundary line).  For these
misclassified type-2 AGN, we find that 38\%, 36\% and 26\% of their
host galaxy emission is dominated by the E, Sbc and Im templates,
respectively, where we define an object to be dominated when $>$50\%
of the host luminosity is coming from a given template.  These
similar proportions suggest that this selection bias, if present,
is not preferentially missing Im galaxies, and therefore it is
unlikely to be significantly affecting our results.  For completeness,
we note that it is difficult to determine the dominant host for the
majority of blue candidates which have $E(B-V)<0.15$ and thus AGN
emission dominates the optical SED, making host SED fitting
challenging.

The same analysis for red AGN candidates in Bo\"{o}tes gives 63\%,
13\% and 24\% of the cases dominated by E, Sbc and Im templates,
respectively. We note here that an E galaxy SED template does not
directly imply that the galaxy morphology is elliptical. As discussed
below, sources with the E-type template also include spiral galaxies
with prominent bulges. Overall, there is also a large fraction of
objects (37\% if we combine Sbc and Im) dominated by late-type
templates, suggesting that while early-type hosts are common, the
red AGN population is hosted by a mixture of galaxy types.

Second, we recall the morphological results from \S\ref{sec:morph}.
There we found that 54\% (20\%) of red (blue) AGN have disks, while
32\% (80\%) are elliptical or point sources. This means that the
red AGN sample is not dominated by typical red sequence galaxies,
and that blue AGN have, in fact, a lower fraction of late-types
than red AGN. These findings
strongly suggest that it is unlikely that the clustering results
are driven by host galaxy differences or selection bias.  Instead,
the observed differences in their correlation functions actually
represents an intrinsic difference in the environments of type-1
and type-2 AGN. Furthermore, as discussed in \S\ref{sec:discussion},
the increase in clustering while moving from blue cloud to red
sequence galaxies is markedly smaller than the difference between
blue and red AGN.

\subsection{Sensitivity to Redshift Distribution}

As the amplitude of $w(\theta)$ will certainly change depending on
the location and shape of the redshift distribution in any observed
sample of galaxies, it is important to assess how sensitive our
bias estimates are to changes in the redshift distribution.

\begin{figure}
\epsscale{1.2}
\plotone{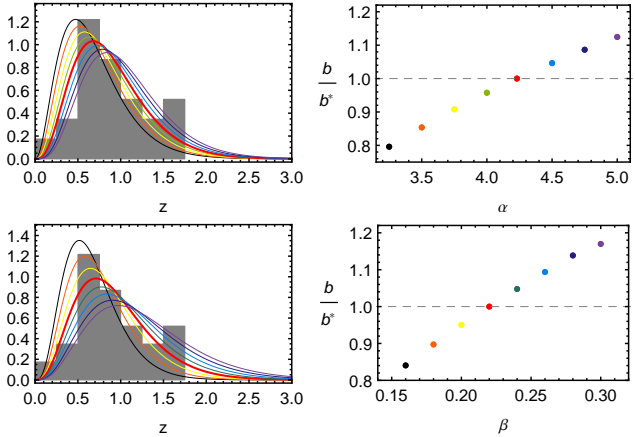}
\caption{{\bf Top row:} model redshift distributions of varying shape ($\alpha$) parameter 
along with the corresponding change in absolute bias normalized to the best-fitting 
case (red thick line) to COSMOS spectroscopic data for red AGN (solid histogram). 
{\bf Bottom row:} same as before but for distributions of varying scale ($\beta$) 
parameter.}
\label{fig:varbias_red}
\end{figure}

\begin{figure}
\epsscale{1.2}
\plotone{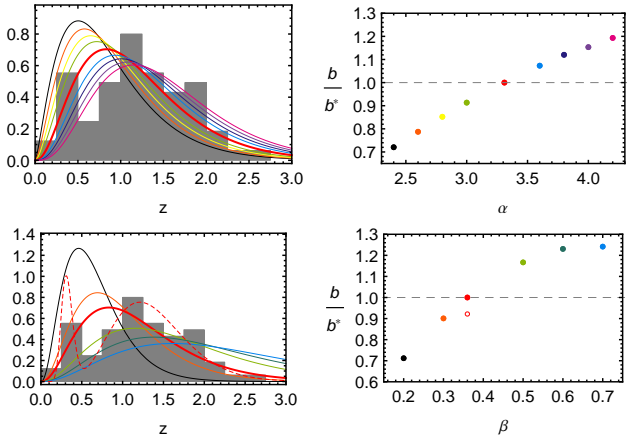}
\caption{Same as Figure \ref{fig:varbias_red}, but for the blue AGN sample. In 
addition, the bottom panels show the fit of a double Gamma distribution (dashed 
red line) along with the change in absolute bias (empty circle) with respect to 
the single distribution case.}
\label{fig:varbias_blue}
\end{figure}

For this purpose we fit different model distributions for red and blue AGN redshifts
and investigate how much the inferred absolute bias would change by systematically 
varying the distribution parameters with respect to the best-fitting case. To describe our 
redshifts, we adopt the Gamma statistical distribution of parameters $\alpha$ (shape) 
and $\beta$ (scale), although this choice is not critical (e.g. using Gaussians will 
lead to variations of the same order of magnitude). For red AGN the best-fit parameters
to observed COSMOS redshifts are $\alpha=3.98$ and $\beta=0.22$ ($\langle z\rangle=0.88$), 
while for blue AGN we obtain $\alpha=2.65$ and $\beta=0.39$ ($\langle z\rangle=1.03$).

For our red AGN sample, in Figure \ref{fig:varbias_red} we reproduce the different model 
distributions with varying $\alpha$ (top row) and $\beta$ (bottom row), and the 
corresponding effect on the bias, always normalized to the best-fitting case 
highlighted in red. Changing $\alpha$ or $\beta$ within the range shown means that
the bias could change by $\sim$20\% at most. Note that to estimate $b/b^*$ we 
assume a spatial correlation length $r_0$ that is constant in redshift.

Figure \ref{fig:varbias_blue} shows the same analysis applied to our blue AGN sample. 
The result is a similar variation of $\sim$25\% in bias. In addition, we also test the 
effect of fitting a double Gamma distribution (dashed red line) instead of a single one. 
As expected, adding a second peak to the model naturally adjusts much better to the 
observed redshifts, yet the derived bias would decrease by only $\sim$6\%. To further 
assess these conclusions, we repeated the test by directly convolving the COSMOS 
redshift distributions of blue and red AGN with a Gaussian kernel of increasing width. 
Once again the bias changes by about 30\% for any reasonable broadening. Note that 
adopting the Bo\"{o}tes redshift distributions as reference instead of COSMOS shifts 
these percentages by $\lesssim$4\%, and so does not qualitatively alter our conclusions.
Finally, a bias uncertainty of $\sim20$\% in the \textit{least favorable} scenario ---
e.g., the blue AGN bias is underestimated by 20\%, while the red AGN bias is 
overestimated by 20\% --- still translates into systematic halo mass estimates a factor 
of $\sim3$ larger for red AGN than for blue AGN. 

\section{Discussion}
\label{sec:discussion}
In this work we have taken advantage of recently released data from
\wise\ to construct a large sample of $\sim 170,000$ mid-IR-selected
AGN candidates with the main purpose of analyzing their angular
clustering properties. The selection is highly reliable ($> 90\%$),
as demonstrated in \citet{stern2012} and \citet{assef2013}, as well
as by the high rate of X-ray detections (\S 3.1). The median redshift
of the sample is $\langle z \rangle \sim 1.1$ based on relatively
complete spectroscopy in the COSMOS and Bo\"{o}tes fields. By
considering their optical counterparts from SDSS, we distinguish
those \wise-selected AGN that are optically faint, and thus have
red optical-to-mid-IR colors and are inferred to be heavily obscured AGN,
from those that are optically bright, and thus have blue optical-to-mid-IR
colors and are inferred to be unobscured AGN.

We find that, as a whole, the \wise-selected AGN population presents
a similar clustering strength to optically selected quasars at
comparable redshifts, with a slightly higher absolute bias with
respect to the dark matter distribution for redder $W1 - W2$
subsamples. We find that the red AGN show a notably larger bias 
level than that of blue AGN, with $b = 2.9 \pm 0.6$ versus 
$b=1.6\pm0.6$ respectively. Using a significantly smaller sample of 
few hundred sources over a much smaller area, \citet{hickox} reported 
a similar absolute bias of $b = 2.87 \pm 0.77$ for obscured 
{\it Spitzer}-selected AGN. Our absolute bias estimates suggest that 
red AGN ({\it i.e} obscured sources) are hosted by massive
dark matter halos of $\log(M/M_\odot\, h^{-1}) \sim 13.5$, 
well above the halos of mass $\log(M/M_\odot\, h^{-1}) \sim 12.4$ 
that harbor blue AGN (unobscured sources). 

It is possible to interpret these results in a scenario where, at least 
during a brief phase before the 
dust is removed and the AGN gets ``exposed'', the black hole mass is a 
factor of few below the $M-\sigma$ relation of active galaxies. For our 
sample, from the SED fits of \wise\ AGN in
Bo\"{o}tes we find that both red and blue AGN have similar distributions
of AGN bolometric luminosity, with a nearly identical mean of $\sim
2 \times 10^{12} L_{\odot}$. This suggests that the black hole
masses of our red and blue AGN do not differ much, and it is unlikely
that their relative Eddington ratio is much different from unity.
Moreover, their high luminosities are indicative of quasar-like
accretion happening in both samples and we know that \wise\ AGN selection 
tends to pick up AGN radiating at large fractions of their Eddington limits
(\citealt{assef2013}). Since we find that obscured sources 
are hosted by more massive halos, then this means that, at least during 
a period of time, the black hole mass growth 
lags behind that of the hosting halos and hence the black holes in 
obscured AGN are temporarily ``undermassive'' until they reach their final mass.
This is not entirely surprising, as, for example, \citet{alexander} find that 
submillimeter galaxies at $z = 2$ host black holes $\sim 10$ times 
smaller than the expected for radio galaxies and quasars.

The basis of such a lag argument for AGN has been proposed before in the
literature (e.g., theoretically by \citealt{king}, and coupled to
clustering by \citealt{hickox}). \citet{king} suggests that the
effect of Rayleigh-Taylor instabilities on the Eddington outflows
that regulate black hole growth leads to black holes masses in
active galaxies a factor a few below the M-$\sigma$ relation,
assuming an observed AGN phase represents a black hole growth phase.
Thus, AGN should recurrently reach Eddington-order luminosities in
order to grow fast enough to reach the masses specified by the
\citet{soltan} relation.

One popular scenario for obscured quasars is that they represent
an early evolutionary stage of rapid black hole growth just before
the emergence of an unobscured, optical quasar. \citet{hopkins08}
pose that in the final stages of coalescence of the galaxies, massive
inflows supply large amounts of gas, increasing the gas density
around nuclear regions and feeding the black hole that:  (1) initially
is obscured, (2) grows accordingly at high Eddington rates, and (3)
is small compared to the spheroid in formation. Then, any possible
link between (final) black hole mass and halo mass (e.g.
\citealt{ferrarese}) would predict DM halos of the same mass for
obscured and unobscured sources. But, if obscured AGN are an early
stage where black holes are acquiring their final mass, then they
would inhabit more massive halos when compared to unobscured quasars
of the same black hole mass. 

The clustering of red and blue galaxies has been studied in detail
by \citet{coil08} using DEEP2 survey data. They find that at $z
\sim 1$ the bias of blue cloud galaxies is in the range of $b \sim
1.2-1.4$. Moving towards the red sequence, the bias increases in
about 30\%, so the measured bias of red galaxies is $b \sim 1.6-1.8$.
Our blue AGN candidates have a bias that is at least comparable to
luminous blue DEEP2 galaxies or to their less luminous red galaxies,
but the bias of \wise\ red AGN is much larger than that of red
galaxies, and is well more than 30\% greater than that of blue AGN.
This suggests that our red AGN candidates do not seem to cluster
like typical red sequence galaxies at these redshifts and that the
change in clustering is intrinsic to the two AGN types. These results
are in broad agreement with \citet{hickox09}, who finds that
mid-IR-selected AGN tend to reside in galaxies slightly bluer than
the green valley; and with \citet{griffith}, who conclude that the
X-ray and mid-IR AGN are not dominated by early-type galaxies, but
by later-type galaxies with disks.

Finally, our results allow to test a basic assumption of the AGN 
unification paradigm. A fundamental prediction of orientation-driven 
AGN unification models is
that the angular clustering strength should be similar for obscured
and unobscured AGN. We find evidence that obscured AGN are, in fact,
more clustered than unobscured sources, which would appear to make
simple orientation, or obscuring torus scenarios much less plausible,
or, at least, not the full story for AGN obscuration.  Alternatively,
it would be interesting to compare our results against predictions
of more physical AGN models, where, for example, the sublimation
radius changes with AGN power or the covering fraction depends on
other physical parameters.  Our data set does not allow us to test
these models in detail, but larger samples with improved redshifts
and spectral coverage will make it possible. Our primary result is
a significant difference in the clustering of optically bright
(blue) and optically faint (red) mid-IR-selected AGN, implying that,
on average, obscured and unobscured AGN reside in different halos.
This surprising result has important implications for AGN unification,
the role of AGN feedback in galaxy formation, the lifetime of
quasars, and for understanding the sources responsible for the
cosmic X-ray background and their cosmic evolution.

\acknowledgments
We thank A. Myers for useful replies to questions and extend our
gratitude to the {\it WISE} extragalactic science team for its
continuous support and interesting discussions over the years.  We
also gratefully acknowlege the anonymous referee and numerous
colleagues including G. Hasinger and R. Hickox who provided insightful
comments that have improved our discussion.  This publication makes
use of data products from the {\it Wide-field Infrared Survey
Explorer}, which is a joint project of the University of California,
Los Angeles, and the Jet Propulsion Laboratory/California Institute
of Technology, funded by the National Aeronautics and Space
Administration. Funding for the SDSS and SDSS-II has been provided
by the Alfred P. Sloan Foundation, the Participating Institutions,
the National Science Foundation, the US Department of Energy, the
National Aeronautics and Space Administration, the Japanese
Monbukagakusho, the Max Planck Society and the Higher Education
Funding Council for England. The SDSS website is {\tt
http://www.sdss.org/}. R.J.A. was supported in part by an appointment
to the NASA Postdoctoral Program at the Jet Propulsion Laboratory,
administered by Oak Ridge Associated Universities through a contract
with NASA.  R.J.A. was also supported in part by Gemini grant number
32120009. We also thank the NASA Astrophysics Data Analysis Program
(ADAP) for its support.

\end{document}